\documentclass[preprint,12pt,authoryear]{elsarticle}

\usepackage[T1]{fontenc}

\usepackage{graphicx}
\graphicspath{ {figures/} }

\usepackage{array}
\usepackage{tabularx}
\usepackage{booktabs}     
\usepackage{xcolor}       

\usepackage{xurl}
\usepackage{hyperref}
\usepackage{capt-of}
\usepackage{placeins}
\hypersetup{hidelinks}

\usepackage[shortcuts]{extdash}
\newcolumntype{L}[1]{>{\raggedright\arraybackslash}p{#1}}
\newcolumntype{Y}{>{\centering\arraybackslash}X}

\journal{Computers \& Security}

\begin{document}

\begin{frontmatter}

\title{APIOT: Autonomous Vulnerability Management Across Bare-Metal Industrial OT Networks}

\author[kent]{Adel ElZemity\corref{cor1}}
\ead{ae455@kent.ac.uk}

\author[kent]{Budi Arief}
\author[kent]{Shujun Li}

\author[bristol]{George Oikonomou\corref{cor1}}
\ead{g.oikonomou@bristol.ac.uk}

\author[bristol]{James Pope}

\cortext[cor1]{Corresponding authors}

\affiliation[kent]{organization={University of Kent},
            city={Canterbury},
            country={United Kingdom}}

\affiliation[bristol]{organization={University of Bristol},
            city={Bristol},
            country={United Kingdom}}

\begin{abstract}
Operational technology (OT) devices run safety-critical physical processes, yet their security testing remains manual and expert-intensive. Autonomous security agents could ease this burden, but bare-metal OT devices are difficult targets because they lack shells, filesystems, and named exploit abstractions, requiring operation through protocol fields and observable device state. We present APIOT (Autonomous Purple-teaming for Industrial OT), a large language model (LLM) agent framework that completes a discovery $\to$ exploitation $\to$ network-level mitigation $\to$ verification workflow without human intervention. We evaluate APIOT in 360 runs on vulnerable firmware and simulators for Zephyr RTOS, spanning five LLMs, three topologies, two impairment levels, and prompt, tooling and two ablations. APIOT completed 90\% of primary missions (63/70 across protocols and topologies) and 85\% of 150 cross-model runs (128/150) --- feasible across models, with substantial variation. Named exploit tools (e.g., a canned CoAP option-overflow) were faster, but protocol primitives (the agent builds the fields itself) produced more reusable mitigations that blocked unseen payload variants. Runtime oversight improved reliability: success rose from 27/30 to 29/30 with rule-based guards and 30/30 when a second LLM added advisory steering. To our knowledge, this is the first end-to-end autonomous attack-remediation demonstration on bare-metal industrial devices, establishing controlled feasibility --- not zero-day discovery or production-OT transfer.
\end{abstract}

\begin{keyword}
OT security \sep IoT security \sep autonomous vulnerability management \sep LLM agents \sep bare-metal MCU \sep industrial protocols
\end{keyword}

\end{frontmatter}

\section{Introduction}

Operational technology (OT) devices, such as industrial controllers, field sensors, and smart meters, present a growing and under-protected attack surface~\citep{bhamare2020cybersecurity,Stouffer2015GuideTI}. Security testing of these systems today is a manual, expert-intensive process: a practitioner must understand the protocol, craft raw payloads, and interpret device responses. This is partly because, unlike IT infrastructure, these devices run bare-metal firmware on resource-constrained microcontrollers (MCUs): at most a minimal real-time operating system rather than a general-purpose one, a rudimentary platform-specific shell typically disabled or stripped in production builds, and only kilobytes rather than megabytes of RAM. They communicate via industrial protocols such as Modbus/TCP~\citep{modbus2012appspec} and CoAP~\citep{shelby2014coaprfc7252}, which carry device-specific semantics that differ fundamentally from HTTP and SSH. The result is both an engineering bottleneck (scarce expert labour) and a research gap on whether autonomous agents can perform protocol-level exploitation and remediation on bare-metal targets, where device diversity and the scarcity of industrial protocol documentation in large language model (LLM) training corpora leave open whether frontier models can reason at the parser level without fine-tuning.

A first research question therefore concerns \emph{capability}: can autonomous LLM agents complete a protocol-level attack-remediation loop on bare-metal MCU targets? Recent research work has explored whether LLMs can automate penetration testing. Researchers behind PentestGPT~\citep{deng2024pentestgpt}, HackingBuddyGPT~\citep{happe2023gettingpwnd}, and AutoAttacker~\citep{xu2024autoattacker} have shown that LLM agents can navigate Linux privilege escalation and web application attacks with notable autonomy. However, published evaluations overwhelmingly target Linux or web systems. To our knowledge, no prior study has demonstrated an autonomous LLM agent completing a full exploit $\to$ network-level mitigation $\to$ verification cycle on bare-metal MCU targets. This remains a difficult research problem because these targets provide no shell or filesystem abstractions, and successful operation requires protocol-level reasoning rather than system-level intuition.

Recent industry efforts corroborate this trend on conventional IT targets. Meta's CyberSecEval~3 benchmark evaluates LLMs on autonomous offensive cyber operations~\citep{wan2024cyberseceval3}, and OpenAI's o1 system card reports steadily improving performance on offensive capture-the-flag evaluations~\citep{openai2024o1systemcard}. Agentic vulnerability research has also produced real-world findings: Google Project Zero and DeepMind's Big Sleep agent discovered a previously unknown, exploitable vulnerability in SQLite in 2024~\citep{google2024bigsleep} and, in 2025, pre-empted a second SQLite vulnerability known only to threat actors (CVE-2025-6965) before it could be exploited in the wild~\citep{google2025bigsleep6965}. Malicious use is following the same curve: OpenAI reports continuing abuse of its models by state-linked and criminal threat actors for cyber operations~\citep{openai2025disrupting}, and Anthropic disrupted the first reported AI-orchestrated cyber espionage campaign (GTG-1002), in which an agent executed the bulk of the intrusion lifecycle autonomously~\citep{anthropic2025espionage}; Anthropic's 2026 analysis of 832 accounts banned for malicious cyber activity found the share of medium- or high-risk AI-enabled actors rising from 33\% to 56\% within a single year~\citep{anthropic2026attacknavigator}. All of these efforts target conventional IT systems---software binaries, web services, and enterprise networks---leaving bare-metal MCU targets and industrial protocols unaddressed.

A second research question concerns \emph{reliability}. LLM agents in autonomous loops are known to exhibit degenerate behaviours: they repeat failed actions, get stuck in subgoal loops, and make inconsistent progress~\citep{liu2024agentbench,yao2023react}. Prior work addresses this primarily through prompt engineering~\citep{wang2023selfconsistency} or agent-level retries~\citep{shinn2023reflexion}. We ask whether an overseer runtime governance layer~\citep{xiang2025guardagent} (rule-based guardrails with optional LLM-based advisory steering) is sufficient to achieve reliable mission completion without fine-tuning, and how large the performance loss is when the overseer is removed.

To address these questions, we designed \textbf{APIOT (Autonomous Purple-teaming for Industrial OT)}, an LLM agent framework that completes the full discovery, exploitation, network-level mitigation, and verification cycle on bare-metal OT targets without per-step human intervention. APIOT gives the agent access to low-level protocol primitives rather than named exploit tools, requiring it to operate over protocol fields and observed device behaviour, and includes an overseer runtime governance layer that enforces reliability through rule-based guardrails with optional LLM advisory steering. We evaluated APIOT in controlled experiments totalling 360 runs---70 primary capability runs plus 40 tool-abstraction, 60 runtime-governance, 40 impairment, and 150 model-sensitivity runs (Section~\ref{sec:method})---spanning five frontier LLMs (MiniMax~M2.5, Gemini~3.1~Pro, Claude~Sonnet~4.6, GPT-5.4, and GLM-5), three network topologies, two impairment levels, and two direct ablations (tool abstraction and runtime governance). The evaluated firmware intentionally exposes known protocol-edge-case weaknesses; our goal is to measure autonomous protocol-level operation and network-level mitigation, not to claim zero-day discovery. Our main contributions are:

\begin{enumerate}

\item \textbf{An empirical demonstration of autonomous attack--remediation on bare-metal MCU targets}: To our knowledge, APIOT is the first end-to-end autonomous LLM agent to complete discovery $\to$ exploit $\to$ network-level mitigation $\to$ verify cycles on bare-metal MCU targets running industrial protocols, without per-step human intervention and across direct and multi-hop OT topologies (Section~\ref{subsec:capability}). The evaluation uses Zephyr RTOS firmware on ARM Cortex-M MCUs and crash-semantics-equivalent simulators; since the exploits target protocol application layers rather than CPU-specific behaviour, the results are not tied to a specific MCU core. A five-model sensitivity study shows the capability is feasible across frontier LLMs, though with substantial model-to-model variation (Section~\ref{subsec:generalisability}).

\item \textbf{A tool-abstraction ablation}: We compare protocol-primitive tools against named exploit tools that trigger the same known weaknesses, testing whether the evidence reflects protocol-level construction rather than tool-name selection (Section~\ref{subsec:primitive_ablation}). Named tools are faster, but primitives yield observable field construction and mitigations that generalise better to payload variants.

\item \textbf{A controlled runtime-governance ablation}: We separate prompt-only operation, deterministic rule guards, and guards plus optional LLM advisory steering (Section~\ref{subsec:oversight}). Deterministic guards account for most of the reliability gain, with advisory steering adding a smaller incremental benefit.

\item \textbf{A taxonomy of systematic failures and a deployment envelope}: Across all observed mission failures, we identify four systematic patterns (protocol confusion, stall-and-repeat loops, phase deadlock, and infrastructure failure), enabling targeted countermeasures. In our experiments, the deployment boundary for autonomous CoAP operation appeared in the three-tier Edge--Fog--Cloud topology~\citep{mohan2016edge}, where traffic traverses multiple hops before reaching the target (Section~\ref{subsec:capability}).

\end{enumerate}

The rest of this paper is organised as follows. Section~\ref{sec:background} provides background and related work. Section~\ref{sec:design} presents the APIOT architecture, and Section~\ref{sec:method} details the experimental methodology. Section~\ref{sec:results} reports empirical results, while Section~\ref{sec:discussion} discusses the implications of these results. Section~\ref{sec:conclusion} concludes the paper and outlines directions for future work. Appendix sections provide implementation and configuration details.

\section{Background and Related Work}
\label{sec:background}

\subsection{Bare-Metal MCU Security and Testbeds}

Resource-constrained MCUs running real-time operating systems (RTOS) such as Zephyr~\citep{zephyrproject2024docs}, FreeRTOS~\citep{freertos2023}, or Contiki-NG~\citep{oikonomou2022contiking} form the lowest layer of the OT stack. A typical Cortex-M3 device~\citep{cortexm3trm} has 64--256\,KB of RAM, 256\,KB of flash, and runs a single firmware image with no MMU, no process isolation, and no dynamic linking. This means exploitation has no shell to land in and no filesystem to pivot through: success is measured by crashing the device (denial of service) or writing to control registers (actuation attack). We focus on the ARM Cortex-M family (M3 and M4F in our experiments) as one of the most widely deployed 32-bit MCU classes in OT devices and the best supported by mature full-system emulation; ESP32-class devices share the same bare-metal constraints, and because the attacks in this study operate at the protocol application layer rather than on CPU-specific behaviour, the choice of core does not materially affect the results. Among the available RTOS options we use Zephyr: it is open source and vendor-neutral, supports the emulated Cortex-M boards in our testbed, and its Kconfig build system allows a minimal, shell-less image matching the constraints above (details in \ref{app:iovlab}).

\textbf{CoAP} (Constrained Application Protocol, RFC~7252; \citealp{shelby2014coaprfc7252}) is a UDP-based request-response protocol designed for resource-constrained nodes. It supports confirmable (CON) and non-confirmable (NON) messages; option fields encode resource paths and metadata. Malformed option headers (delta/length overflow) can exhaust parser state on devices with no bounds checking~\citep{Alghamdi2013SecurityAO,Arvind2019AnOO}.

\textbf{Modbus/TCP} maps the Modbus application layer over TCP port~502. It uses a Modbus Application Protocol (MBAP) header carrying a length field; devices trusting this field without validation are vulnerable to heap overflow~\citep{Lai2020VulnerabilityMM} when the declared length far exceeds the actual payload.

\textbf{MQTT}~\citep{oasis2014mqtt311} is a publish--subscribe protocol (TCP/1883) widely deployed for IoT telemetry. Nodes publish messages to named topics on a central broker and subscribe to topics of interest; brokers without enforced access control allow any client to read wildcard topic filters or write to command topics, exposing actuation channels to unauthorised publishers. These three protocols were chosen to cover the two dominant OT interaction styles: request--response access to device state (CoAP, Modbus/TCP) and brokered publish--subscribe telemetry and command (MQTT).

\subsection{Autonomous Penetration Testing and Agent Reliability}

\subsubsection{Autonomous exploitation of conventional IT targets}

\citet{deng2024pentestgpt} used GPT-4 in a reasoning-generation-tool loop to guide human-driven penetration tests, achieving competitive capture the flag (CTF) performance. \citet{happe2023gettingpwnd} demonstrated fully autonomous Linux privilege escalation with HackingBuddyGPT via tool-calling LLMs. \citet{xu2024autoattacker} extended Happe and Cito's work to network reconnaissance and lateral movement on Linux hosts with AutoAttacker. \citet{Happe2023LLMsAH} systematically evaluated LLMs' capabilities for privilege escalation across multiple models. \citet{Fang2024LLMAC} showed that GPT-4 agents could autonomously exploit 87\% of one-day Common Vulnerabilities and Exposures (CVEs) given the vulnerability descriptions, extending autonomous exploitation to real-world CVE databases.

All of the above studies evaluated Linux targets with full POSIX semantics. To the best of our knowledge, prior work has not considered bare-metal MCU targets, industrial protocols, or the or the two-phase red-then-blue workflow~\citep{vanbuggenhout2024purple,scarfone2008nist800115}. Our work addresses this research gap.

\subsubsection{Agent reliability and runtime governance}

\citet{yao2023react} introduced the thought-action-observation loop as a standard LLM agent architecture. \citet{liu2024agentbench} showed that 29 API-based and open-sourced LLMs struggle with long-horizon task completion, exhibiting repetition and inconsistency in complex agentic settings. Recent work on LLM agent safety and security has proposed various monitoring strategies, including prompt-level instructions and secondary LLM judges~\citep{DasSecurity,li2025securityconcerns,mou2026toolsafeenhancingtoolinvocation,raptis2025agenticrobotics,wang2025agentspeccustomizableruntimeenforcement}. This literature leaves open whether lightweight rule-based guardrails, optionally augmented by a secondary LLM for advisory steering rather than as the primary enforcement mechanism, are sufficient to prevent dominant failure modes.

\subsubsection{Firmware testbeds and discovery on bare-metal targets}

\citet{kim2020firmae} and \citet{chen2016firmadyne} demonstrated that emulation of real firmware can be scaled to thousands of Linux-based router images, establishing dynamic analysis of unmodified firmware as a practical evaluation substrate. Their systems, however, serve automated analysis pipelines rather than interactive agent workloads, and neither emulates bare-metal MCUs. \citet{Ghazanfar2020IoTFlockAO} proposed IoT-Flock, an open-source framework for generating realistic IoT protocol traffic, which is useful for testing detection systems but models neither bare-metal devices nor adversarial agent workflows. Together, these works provide the emulation and traffic-generation foundations on which we build, while leaving open the need for a testbed that boots bare-metal MCU firmware alongside Linux devices and exposes it to an autonomous agent---the role of IoT Virtual Lab in this paper.

For the discovery stage, the most directly comparable prior work on bare-metal MCUs is firmware fuzzing (automated generation of malformed inputs to discover crashes and memory corruptions in embedded firmware). \citet{HALucinator}, \citet{Fuzzware}, and \citet{P2IM} emulated MCU firmware and applied coverage-guided fuzzing to find crashes automatically; unlike our framework, this line of work does not address autonomous exploitation, mitigation, or verification stages. APIOT is complementary rather than competing: it operates as a reasoning agent (not a fuzzer), targets the full discovery $\to$ exploitation $\to$ network-level mitigation $\to$ verification cycle, and requires semantic understanding of \emph{why} a payload triggers a parser failure rather than discovering it by exhaustive mutations. IoT Virtual Lab, our QEMU-based emulation testbed, differs in supporting Cortex-M3 bare-metal firmware alongside Linux devices on a shared network, with RESTful API control, multiple industrial IoT (IIoT) topology models, and configurable network realism, properties designed to support autonomous agent evaluation.

Our experiments deliberately use known or seeded protocol-edge weaknesses in CoAP, Modbus/TCP, and MQTT. This choice makes the study a controlled evaluation of autonomous protocol-level operation, not a zero-day discovery claim. The question is whether an LLM agent, given low-level protocol and mitigation primitives, can discover the relevant target state, construct payloads, verify impact, derive a network-level mitigation, and validate that mitigation without step-by-step human control. Authenticated and encrypted variants, including CoAP over DTLS, MQTT with enforced broker ACLs/TLS, Modbus security gateways, OPC UA, and DNP3 deployments, are therefore outside the evaluated scope and motivate the secured-protocol experiments discussed in Section~\ref{sec:discussion}.

\begin{figure}[t]
\centering
\includegraphics[width=\linewidth]{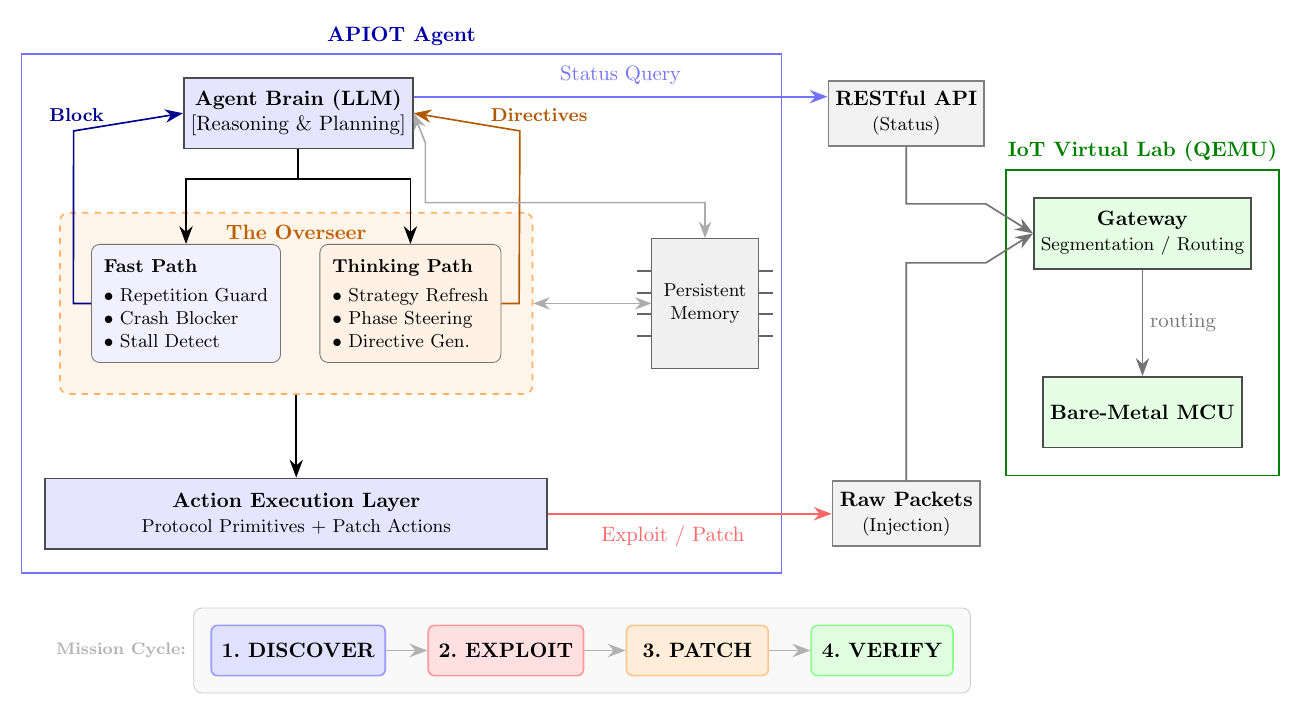}
\caption{High-level architecture of the APIOT autonomous purple-teaming framework and its interface with the IoT Virtual Lab testbed (see Section~\ref{sec:design} for details).} 
\label{fig:arch}
\end{figure}

\section{System Design}
\label{sec:design}

Figure~\ref{fig:arch} shows the overall architecture of our framework. APIOT is built around an Agent Brain, an LLM reasoning loop backed by a persistent memory store, that drives a four-stage mission cycle of discovery, exploitation, network-level mitigation, and verification. An Overseer sits between the Agent Brain and the Action Execution Layer and enforces reliability through two paths: a rule-based fast path (repetition and crashed-target guards, phase-transition enforcement) and an optional LLM-based thinking path that provides advisory steering when the fast path detects a stall. IoT Virtual Lab emulates the target network, and APIOT interacts with it solely through low-level raw packets and a read-only RESTful API for status queries; APIOT has no lifecycle control over IoT Virtual Lab, which ensures that measured capability reflects in-mission reasoning rather than orchestration privileges.

\subsection{Design Objectives and Scope}

The design of APIOT is driven by four requirements. \textbf{(R1) End-to-end autonomy}: the agent must complete discovery $\to$ exploitation $\to$ network-level mitigation $\to$ verification within one mission, without per-step operator control. \textbf{(R2) Protocol-level reasoning}: the capability test should measure if the agent can infer exploit semantics from protocol structures, not retrieve canned exploit names. \textbf{(R3) Runtime reliability}: the architecture must prevent common long-horizon agent failures (repetition, stalled reconnaissance, and invalid phase progression). \textbf{(R4) Experimental isolation and reproducibility}: the target environment must be resettable, observable, and strictly separated from agent lifecycle control.

\subsection{Core Design Decisions}

\textbf{Protocol primitives over named exploits.} The action space exposes low-level protocol primitives rather than pre-packaged exploit labels. This is intended to require explicit selection of protocol fields and parser-facing inputs rather than merely selecting a canned exploit name. Because tool schemas themselves provide affordances, this design choice is evaluated directly through a named-exploit-tool ablation in Section~\ref{subsec:primitive_ablation}.

\textbf{Two-phase purple-team workflow.} The mission is structured as Red (discover/exploit/confirm impact) followed by Blue (derive and verify a network-level mitigation). This decomposition aligns the objective with operational purple-team practice~\citep{vanbuggenhout2024purple}, in which offensive findings are converted directly into validated defensive improvements, and makes failure attribution clearer than a monolithic task objective.

\textbf{Runtime governance.} Overseer places rule-based guardrails on tool-call sequences and mission progression, with optional LLM advisory steering only as a supplementary mechanism. This separates structural safety guarantees from model quality and enables direct ablation of the governance premise.

\textbf{Strict environment boundary.} The agent can observe topology and send packets, but it cannot spawn, reset, or terminate lab devices. This avoids hidden orchestration privileges and ensures measured capability comes from in-mission reasoning and action selection.

\subsection{Agent Harness and Prompt Conditions}

The harness is part of the method: APIOT implements a continuous thought $\to$ action $\to$ observation loop using an OpenAI-compatible chat-completion client and JSON-schema tool calling. At each turn, the model receives the conversation state and the tool schemas, selects one or more tool calls, and the dispatcher executes the corresponding Python implementation. A mission terminates only when the agent emits \texttt{TASK\_COMPLETE} after both red and blue phases, or \texttt{TASK\_ABORTED} after an unrecoverable configuration, infrastructure, or budget failure.

Two prompt families are used. Both conditions share the same red-then-blue mission rules, the same mandatory crash and mitigation verification requirements, the same 21 JSON-schema tools, the same overseer settings, the same model temperature, and the same termination rules. The guided prompt additionally provides protocol-specific examples for CoAP option encoding, Modbus MBAP length fields, and MQTT command-topic discovery. The blind prompt removes those protocol-specific exploit hints while preserving the same autonomy, safety, target-knowledge, and tool-use rules, leaving the agent to infer the attack from service observations and tool schemas.

\ref{app:artifact_transparency} records the prompt source identifiers, runner flag for blind mode, tool-schema source file, compaction implementation, and overseer implementation used in the reported experiments. This makes the guided/blind contrast reproducible without relying on prose descriptions alone.

\textbf{Context management.} APIOT estimates context use at approximately four characters per token. When a run exceeds 70\% of the selected model's context window, the compaction layer preserves the system prompt and the latest eight messages, replaces the older middle history with a structured summary, and truncates large tool outputs to 2,000 characters. This preserves recent low-level observations while bounding runaway context growth. The trade-off is explicit: old raw tool outputs may be summarised away, so long-horizon decisions may depend on summary fidelity and persistent memory rather than the full original transcript.

\subsection{Protocol-Primitive Tool Interface}

The agent does not call a tool named after a vulnerability; it chooses protocol fields and mitigation signatures. Representative red-team primitives include \texttt{coap\_send}, which takes CoAP message fields plus raw \texttt{options\_hex} and returns response bytes, timeout status, and round-trip time; \texttt{modbus\_request}, which takes a function code, PDU bytes, unit ID, and optional MBAP \texttt{claimed\_length}; and the MQTT pair \texttt{mqtt\_subscribe}/\texttt{mqtt\_publish}, which exposes topic filters, topic names, payloads, and message observations. The verification and blue-team side is similarly primitive: \texttt{verify\_crash} reports alive/crashed state, \texttt{iptables\_rule} applies protocol/port/payload-match rules, and \texttt{verify\_patch} replays the prior payload to test whether the mitigation blocks the previously successful attack.

To separate protocol-level operation from tool-name selection, we additionally define a named-exploit-tool control condition. In that condition, the agent receives high-level red-team tools such as \texttt{coap\_option\_overflow} and \texttt{modbus\_mbap\_overflow}, which execute the same known vulnerability triggers without requiring the model to construct CoAP option bytes or MBAP length fields. The blue-team side remains constrained to network-level mitigation and replay verification. This control tests whether faster original-payload triggering also produces reusable mitigation signatures for payload variants that are not exactly the canned trigger.

\subsection{Overseer Runtime Governance}

The overseer observes each proposed tool call before dispatch. Its fast path is rule-based: it blocks identical calls after three repeats in a 20-call window, prevents attacks against already crashed devices, prevents re-attacking a vulnerability with a verified mitigation, detects stalls after eight turns without progress, and enforces phase transitions when red-team conditions are satisfied. A secondary LLM advisory path can receive a compact state snapshot (phase, recent calls, failures, target status, and untried attacks) and return two or three textual steering directives; it never calls tools directly. We isolate these paths through three execution modes: \texttt{off} bypasses both paths, \texttt{guards-only} retains deterministic enforcement and heuristic fallbacks but disables secondary-model calls, and \texttt{full} enables both guards and advisory steering. This design attributes differences between \texttt{off} and \texttt{guards-only} to deterministic governance, and differences between \texttt{guards-only} and \texttt{full} to the incremental advisory path within the evaluated T1 conditions.

\subsection{Instantiation for This Study}

This conceptual design was instantiated and evaluated on IoT Virtual Lab across CoAP, Modbus/TCP, and MQTT conditions, with topologies T1--T3 (flat star, Purdue Model-segmented, and Edge--Fog--Cloud respectively). Implementation\-/level parameters (e.g., concrete tool roster, guard thresholds, and device/network configuration) are reported separately in \ref{app:apiot}--\ref{app:iovlab}, while the evaluation protocol is defined in Section~\ref{sec:method}.

\section{Experimental Methodology}
\label{sec:method}

The evaluation is structured around our two key claims:
\begin{itemize}
\item \textbf{Claim~1 (capability and robustness)}: given protocol-primitive tooling, an autonomous LLM agent can complete the full attack--remediation cycle on bare-metal MCU targets. This is assessed across protocols, topologies, impairment levels, and models, which also establishes the sensitivity of the capability to prompt condition, topology, impairment, and model choice.

\item \textbf{Claim~2 (oversight as a reliability mechanism)}: a runtime governance layer---deterministic guards, optionally augmented by advisory LLM steering---improves mission reliability over prompt-only operation. This is assessed via a three-arm controlled ablation comparing mission reliability and systematic behavioural failures.

\end{itemize}

Agent decision-making is characterised as a secondary evaluation goal: tool-call distributions and failure modes are logged to describe \emph{how} the agent operates, independently of success/failure outcomes.

\subsection{Experimental Conditions}

Table~\ref{tab:conditions} summarises the experimental conditions. All runs used a full purple-team mission (both red and blue phases mandatory). Each condition was replicated ten times; the run order was randomised. This design yielded 170 core runs ($\approx$85 compute hours; $\approx$30 minutes/run), giving per-cell 95\% Wilson confidence intervals with half-width below 15 percentage points near 100\% success, which is sufficient to distinguish the ablation and topology effects reported in Section~\ref{sec:results}. The primary-condition runs fixed a single backbone model (MiniMax~M2.5 via OpenRouter) to isolate protocol, topology, impairment, and oversight effects; cross-model robustness was evaluated separately in a 150-run model-sensitivity study. The sampling temperature was fixed at the default ($T=0.7$) for all runs. Heavy impairment included background human-machine interface (HMI) traffic. The T1 flat-star condition is the direct attacker-to-device baseline: for CoAP and Modbus it exposes a single reachable MCU target without an intermediate gateway; T2 and T3 add routing and lateral-movement structure.

\begin{table}[!t]
\caption{Experimental conditions. T1\,=\,flat star; T2\,=\,Purdue Model (PERA)-segmented~\citep{Williams1992ThePE}; T3\,=\,Edge-Fog-Cloud.}
\label{tab:conditions}
\centering
\begin{tabular}{@{}L{0.21\linewidth}L{0.75\linewidth}@{}}
\toprule
\textbf{Dimension} & \textbf{Values}\\
\midrule
Protocol & CoAP (UDP/5683), Modbus/TCP (TCP/502), MQTT (TCP/1883)\\
Topology & T1, T2, T3\\
Overseer & Full (guards + advisory), guards only, disabled\\
Impairment & None; Medium (5\% loss, 50\,ms latency, 10\,ms jitter); Heavy (20\% loss, 200\,ms latency, 40\,ms jitter, HMI traffic)\\
\bottomrule
\end{tabular}
\end{table}

\begin{table}[!t]
\caption{Primary evaluation metrics.}
\label{tab:metrics}
\centering
\small
\begin{tabular}{@{}L{0.31\linewidth}L{0.53\linewidth}c@{}}
\toprule
\textbf{Metric} & \textbf{Definition} & \textbf{Claim}\\
\midrule
Mission success rate & Percentage of runs reaching \texttt{TASK\_COMPLETE} & C1\\
Exploit confirmation rate & Percentage of attack attempts followed by confirmed impact & C1\\
Mitigation verification rate & Percentage of network-level mitigations confirmed by \texttt{verify\_patch} & C1\\
Turns to first exploit & Turns before first successful exploit & C1, C2\\
Turns to completion & Total turns until \texttt{TASK\_COMPLETE} & C1, C2\\
Redundant call rate & Repeated (tool,\,ip,\,args) / total calls & C2\\
Stall events & Turns with no new finding or mitigation & C2\\
Overseer intervention rate & Steering messages / total turns & C2\\
Time-to-remediate & Seconds: first exploit $\to$ verified mitigation & C1\\
Token use & Estimated maximum context tokens per run where logged & C1, C2\\
\bottomrule
\end{tabular}
\end{table}

\textbf{Claim~1 evaluation}: 70 runs (CoAP/Modbus $\times$ T1/T2/T3 $\times$ 10 replicates + MQTT T1 $\times$ 10). \textbf{Tool-abstraction ablation}: 40 runs (CoAP/Modbus $\times$ primitive/named-exploit tools $\times$ 10 replicates). \textbf{Claim~2 ablation}: 60 new runs (30 overseer-OFF and 30 guards-only; CoAP/Modbus/MQTT $\times$ T1 $\times$ 10 per arm); the full-overseer arm reuses the corresponding 30 T1 runs from Claim~1. \textbf{Robustness}: 40 runs (2 impairment levels $\times$ 2 protocols $\times$ 10). \textbf{Model sensitivity}: 150 runs (five models $\times$ three conditions $\times$ ten replicates). \textbf{Total: 360 unique runs.} Unless explicitly stated, percentages are reported within their evaluation family rather than pooled across all families.

Six targeted additions were made to support this evaluation, including an overseer-OFF flag, software MCU simulators, a parameterised experiment runner, enriched per-turn logging, analysis scripts, and impairment integration (details in Appendices~\ref{app:apiot}--\ref{app:iovlab}).

\subsection{Metrics}

Table~\ref{tab:metrics} lists the primary metrics derived from the tool call log.
For the global behavioural characterisation, we define four failure categories coded from session logs: \textbf{Protocol confusion} (wrong exploits for device type); \textbf{Stall-and-repeat} (identical (tool, ip, args) tuples called more than three times with no progress); \textbf{Phase deadlock} (red phase completed, blue phase never initiated despite open findings); and \textbf{Infrastructure failure} (device crashed mid-mission causing an inconsistent network state, ending in mission abort).

\section{Results}
\label{sec:results}

The following results address the two key claims in turn. Sections~\ref{subsec:capability}, \ref{subsec:primitive_ablation}, and \ref{subsec:generalisability} evaluate Claim~1: whether protocol-primitive tooling makes bare-metal MCU exploit-remediation tractable in the evaluated setting, whether that effect is distinguishable from named exploit tool selection, and how the capability varies across models, topologies, and network conditions. Section~\ref{subsec:oversight} evaluates Claim~2: whether the runtime governance layer improves reliability relative to prompt-only operation.

\begin{table}[!t]
\caption{Mission success rate by protocol and topology ($n=10$ per cell) for CoAP and Modbus conditions. Median turns and mission duration are for successful runs only.}
\label{tab:capability}
\centering
\small
\begin{tabularx}{\linewidth}{@{}l l Y Y Y@{}}
\toprule
\textbf{Protocol} & \textbf{Topology} & \textbf{Success} & \textbf{Median turns} & \textbf{Median duration}\\
\midrule
CoAP   & T1 & 10/10 (100\%) & 11.0 & 3.0\,min\\
CoAP   & T2 & 10/10 (100\%) & 30.0 & 8.5\,min\\
CoAP   & T3 & 3/10 (30\%)   & 18.0 & 5.0\,min\\
Modbus & T1 & 10/10 (100\%) & 24.0 & 7.5\,min\\
Modbus & T2 & 10/10 (100\%) & 23.0 & 4.5\,min\\
Modbus & T3 & 10/10 (100\%) & 27.0 & 7.5\,min\\
\midrule
\multicolumn{2}{l}{\textbf{Overall}} & \textbf{53/60 (88.3\%)} & -- & --\\
\bottomrule
\end{tabularx}
\end{table}

\subsection{Autonomous Exploitation and Remediation Capability}
\label{subsec:capability}

Table~\ref{tab:capability} summarises the topology matrix outcomes for CoAP and Modbus (60 runs: T1/T2/T3 $\times$ 10 replicates each). The results show an all-but-one pattern: APIOT achieved \textbf{100\%} mission success in every matrix condition except CoAP on three-tier topology (T3), where it achieved 3/10 (30\%). The additional MQTT baseline condition at T1 achieved 10/10 success; including it yields 63/70 (90.0\%; 95\% Wilson confidence interval: 80.8--95.1\%) overall.

The seven failures occurred exclusively on CoAP~T3, the three-tier Edge-Fog-Cloud topology where the MCU target sits behind two forwarding hops on the isolated OT segment. Session log analysis revealed the dominant failure mode as a \textbf{multi-target reasoning stall}: after crashing the first device in the OT cell, the agent generated three consecutive empty LLM responses when attempting to reason about lateral movement to the second hop, triggering the empty-response abort guard. The agent initiated the first exploit correctly within 3--4 turns in all failing runs; the breakdown occurs specifically at the multi-hop lateral movement step, not during initial target identification or exploitation. Modbus succeeded in all topologies including T3, where TCP's stateful connection model provides implicit routing feedback that the stateless CoAP/UDP agent lacks.

\begin{figure}[t]
\centering
\includegraphics[width=0.9\linewidth]{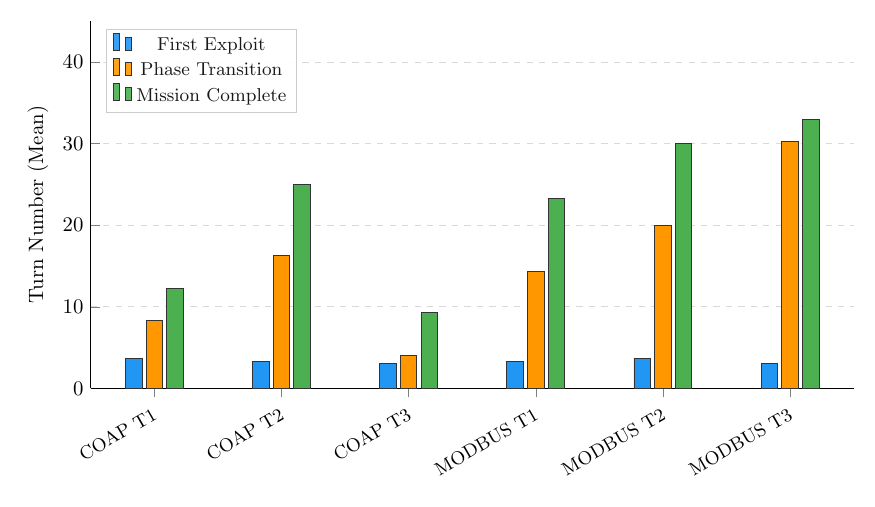}
\caption{Mean turn counts to three milestones: the first exploit attempt, the red-to-blue phase transition (attack complete, defence begins), and the mission completion. The recon-to-exploit interval (turns 3--4) is consistently short, and most of the mission is spent in the exploit-and-verify loop rather than the blue-team phase.}
\label{fig:timeline}
\end{figure}

\textbf{Speed of target engagement.} Across all 60 tabled runs, the agent initiated its first exploit attempt within 5 tool calls, regardless of topology or protocol. This short recon-to-exploit interval shows that, once given protocol-primitive tools and a reachable target, the agent can move from network state to targeted probing quickly without per-step human direction. Figure~\ref{fig:timeline} shows the full timeline distribution.

\paragraph{Agent decision profile.}

\begin{figure}[!t]
\centering
\includegraphics[width=0.9\linewidth]{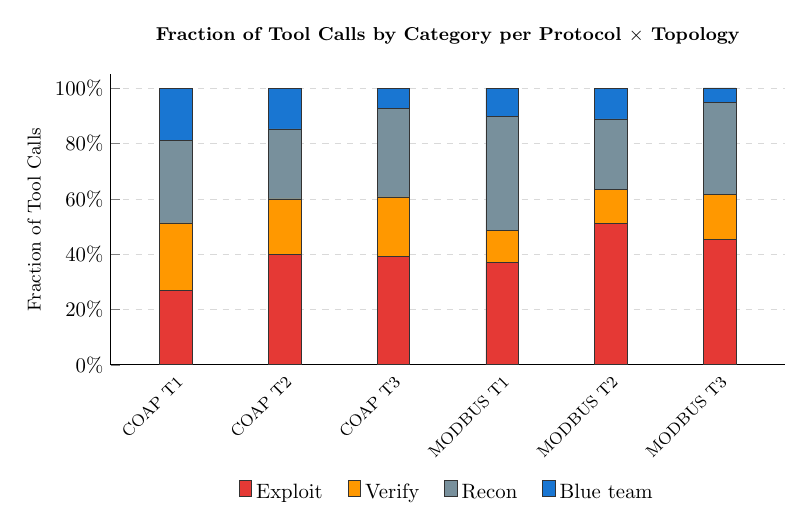}
\caption{Fraction of tool calls by category (exploit, verify, recon, blue team) per protocol $\times$ topology condition. Exploit and recon together dominate in all conditions; blue-team calls are consistently the smallest category (5.0--19.0\%).}
\label{fig:tool_dist}
\end{figure}

Across 70 Claim~1 runs (60 CoAP/Modbus + 10 MQTT), the agent made 1,660 tool calls in total (mean 23.7 per run). Figure~\ref{fig:tool_dist} shows the per-condition tool-call distribution for CoAP and Modbus; MQTT's multi-step attack chain produced a distinct profile discussed in Section~\ref{subsec:oversight}.

\begin{figure}[!t]
\centering
\includegraphics[width=0.71\linewidth]{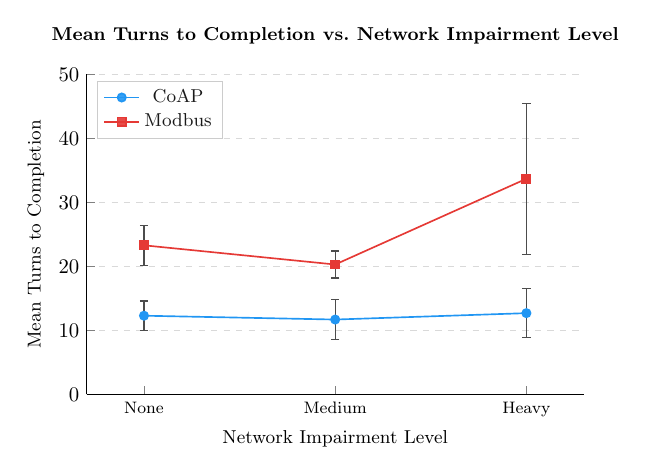}
\caption{Mean turns to mission completion vs.\ network impairment level (all runs successful at 100\%). CoAP turn count is stable across all conditions ($\approx$12 turns); Modbus increases from 23 turns at baseline to 34 turns under heavy impairment (20\% loss, 200\,ms latency), indicating efficiency degradation without mission failure.}
\label{fig:impairment_curve}
\end{figure}

\textbf{Exploit calls} account for 27.0--51.0\% of all calls across conditions (higher in Modbus, where target identification is faster). \textbf{Recon calls} account for 25.0--42.0\%, with CoAP T3 showing the highest fraction (32.0\%), consistent with the extra routing hops requiring more topology discovery. \textbf{Blue-team calls} account for 5.0--19.0\%, a notably small fraction reflecting that remediation is concentrated in a short phase after exploit success.

\textbf{Failure mode.} The seven mission failures (CoAP T3) reflect a lateral-movement planning failure mode specific to stateless UDP protocols in multi-hop topologies, not a target-identification or exploit-generation failure. Zero runs exhibited the ``zero exploits attempted'' or ``exploit succeeded, verify failed'' failure modes, confirming that the agent's protocol-specific exploit logic is reliable once it engages a target. All 63/70 successful runs also completed the blue-team phase, with the mitigation verification step confirmed in every case.

\textbf{Representative attack-remediation traces.} Successful traces illustrate the level at which the agent operates: it constructs protocol fields and network-level mitigation signatures, rather than selecting a pre-named exploit and mitigation pair.
\begin{itemize}
\item \textbf{CoAP.} The agent sent a normal GET (\texttt{40011234}), then used \texttt{coap\_send} with malformed option bytes \texttt{DD FF FF}. The target timed out; \texttt{verify\_crash} confirmed that the CoAP port was unresponsive. The mitigation was an \texttt{iptables} hex-string drop for \texttt{DD FF FF} on UDP/5683, followed by replay verification with \texttt{verify\_patch}.
\item \textbf{Modbus.} The agent started with an FC3 read probe, then tested anomalous MBAP \texttt{claimed\_length} values such as \texttt{0x0800} and \texttt{0x1000}, along with FC5/FC6 write probes. The observed effect was a connection reset or timeout followed by crash/alive verification. The mitigation used \texttt{u32} function-code filters and TCP/502 signature rules, followed by replay verification.
\item \textbf{MQTT.} The agent subscribed to \texttt{\#}, discovered \texttt{iot/sensors/*}, and published \texttt{\{"action":"shutdown"\}} to \texttt{iot/commands/shutdown}. A subsequent subscription observed zero telemetry messages, indicating that the command had taken effect. The mitigation used TCP/1883 string-match rules for shutdown/reboot command payloads or broker ACL hardening.
\end{itemize}

\subsection{Protocol Primitives Versus Named Exploit Tools}
\label{subsec:primitive_ablation}

To test whether the core capability is more than selecting a pre-labelled exploit tool, we compare the protocol-primitive interface against a named-exploit-tool control on the direct T1 baseline. The control exposes high-level tools that trigger the known CoAP option-overflow and Modbus MBAP-length weaknesses directly, while preserving the same mission lifecycle, target discovery, overseer settings, network-level mitigation tools, and replay-verification requirement.

\begin{table}[!t]
\caption{Tool-abstraction ablation on T1 conditions. ``Variant blocked'' measures whether the final network-level mitigation also blocks a held-out payload variant rather than only the original exploit replay.}
\label{tab:primitive_ablation}
\centering
\small
\begin{tabularx}{\linewidth}{@{}l l Y Y Y Y@{}}
\toprule
\textbf{Protocol} & \textbf{Tool interface} & \textbf{Mission success} & \textbf{Median turns} & \textbf{Field construction} & \textbf{Variant blocked}\\
\midrule
CoAP & Primitive & 10/10 & 11.0 & 10/10 & 9/10\\
CoAP & Named exploit & 10/10 & 6.0 & 0/10 & 3/10\\
Modbus & Primitive & 10/10 & 24.0 & 10/10 & 9/10\\
Modbus & Named exploit & 10/10 & 8.0 & 0/10 & 4/10\\
\bottomrule
\end{tabularx}
\end{table}

The ablation is not intended to show that primitives are faster or more successful on the easiest known-trigger setting. Named exploit tools reached impact in fewer turns because they collapse protocol construction into one action. The stronger test is whether the primitive interface produces evidence of protocol-level operation: explicit construction of option bytes or MBAP length fields, derivation of protocol-aware network-level mitigation signatures, and blocking of held-out payload variants. The observed pattern is therefore that named tools are faster on original payloads, but primitives produce higher field-construction and variant-blocking rates.

\textbf{Remediation semantics.} A network-level mitigation in this study is a verified traffic-control rule, not a firmware source-code repair. The agent derives a traffic signature from the crash-inducing payload, applies an iptables/ACL-style rule, and replays the same payload after the target watchdog/reset interval. In the retained remediation artifacts, the applied mitigations are primarily hex-string filters (84 rules), protocol/port drops (80 rules), and Modbus u32/function-code filters (15 rules). This captures a realistic emergency containment action for OT networks, but it does not replace vendor firmware repair or production change-control validation.

\textbf{Token usage.} Token accounting was added during the logging-enrichment phase, so it is available for a subset of retained run databases rather than all evaluated runs. Among 61 runs with positive per-turn token logs, the estimated maximum context use per run ranged from 5,980 to 118,656 tokens; the quartiles were 10,053, 14,767, and 20,872 tokens, with mean 17,705. These values are approximate (character-count based) and should be interpreted as operational-cost indicators rather than provider billing records. Three model-sensitivity runs hit explicit cumulative token-budget guards, which prevented runaway cost in long MQTT/blind failures.

\subsection{Capability Across Topologies, Impairments, and Models}
\label{subsec:generalisability}

Figure~\ref{fig:impairment_curve} characterises the impairment aspect of the autonomous OT agent capability envelope.

\textbf{Topology.} Modbus/TCP achieved 100\% success across all three topologies, including T3. CoAP/UDP succeeded at 100\% on T1 and T2, but degraded to 30\% on T3 via the lateral movement stall and TCP/UDP routing asymmetry described in Section~\ref{subsec:capability}. This establishes three-tier isolated OT cell topology as the current lateral-movement planning boundary for autonomous CoAP deployment, not a network transport limitation.

\textbf{Network impairment.} Both protocols proved robust across all tested impairment levels: CoAP and Modbus each achieved 100\% mission success at medium (5\% loss, 50\,ms latency, 10\,ms jitter) and heavy (20\% loss, 200\,ms latency, 40\,ms jitter, HMI background traffic) conditions. CoAP's UDP-based retransmission handles packet loss without connection state; Modbus/TCP tolerates the latency and jitter introduced without requiring connection re-establishment in these experiments. Network impairment is therefore not a deployment boundary within the tested parameter range.

\begin{figure}[!t]
\centering
\includegraphics[width=\linewidth]{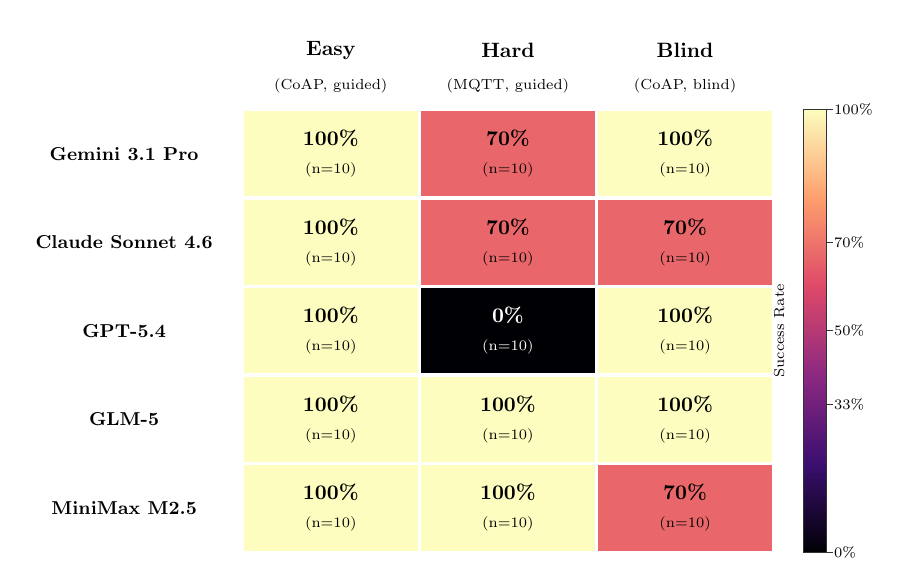}
\caption{Model sensitivity in mission success rate by model and condition ($n=10$ per cell). All models achieve 100\% on easy (CoAP, guided); the hard condition (MQTT, multi-step) separates models, with GPT-5.4 failing all ten runs; the blind condition (CoAP, no protocol hints) shows Claude and MiniMax~M2.5 as models below 100\%.}
\label{fig:rms_heatmap}
\end{figure}

Model sensitivity tests whether the capability demonstrated in Section~\ref{subsec:capability} is tied to one LLM choice. We conducted 150 experiment runs comparing five frontier LLMs (MiniMax~M2.5, Gemini~3.1~Pro, Claude~Sonnet~4.6, GPT-5.4, GLM-5) across three conditions: guided CoAP T1 (\emph{easy}), multi-step guided MQTT T1 (\emph{hard}), and CoAP T1 without protocol hints (\emph{blind}). Figure~\ref{fig:rms_heatmap} summarises the results, while Table~\ref{tab:rms_efficiency} shows the efficiency metrics. The results support cross-model feasibility in the easy condition, but also show that harder multi-step tasks expose large model-specific differences.

\begin{table}[!t]
\caption{Model sensitivity by condition. Turns and redundancy are means over successful runs; dashes denote a cell with no successful runs.}
\label{tab:rms_efficiency}
\centering
\small
\setlength{\tabcolsep}{5pt}
\renewcommand{\arraystretch}{1.04}
\begin{tabular*}{\linewidth}{@{\extracolsep{\fill}}llcrr@{}}
\toprule
\textbf{Model} & \textbf{Condition} & \textbf{Success} & \textbf{Turns} & \textbf{Redundancy}\\
\midrule
Gemini 3.1 Pro     & Easy  & 10/10 & 7.7  & 0.48\\
                   & Hard  & 7/10  & 14.5 & 0.29\\
                   & Blind & 10/10 & 10.3 & 0.43\\
\addlinespace[2pt]
Claude Sonnet 4.6  & Easy  & 10/10 & 16.0 & 0.31\\
                   & Hard  & 7/10  & 31.0 & 0.27\\
                   & Blind & 7/10  & 15.5 & 0.36\\
\addlinespace[2pt]
GPT-5.4            & Easy  & 10/10 & 13.3 & 0.32\\
                   & Hard  & 0/10  & --   & --\\
                   & Blind & 10/10 & 10.3 & 0.34\\
\addlinespace[2pt]
GLM-5              & Easy  & 10/10 & 15.0 & 0.41\\
                   & Hard  & 10/10 & 23.3 & 0.21\\
                   & Blind & 10/10 & 13.3 & 0.45\\
\addlinespace[2pt]
MiniMax M2.5       & Easy  & 10/10 & 12.3 & 0.31\\
                   & Hard  & 10/10 & 21.0 & 0.25\\
                   & Blind & 7/10  & 9.0  & 0.35\\
\bottomrule
\end{tabular*}
\end{table}

\textbf{Easy condition (CoAP T1, guided).} All five models achieved 100\% success. Efficiency diverges: Gemini~3.1~Pro is fastest (7.7~turns), followed by MiniMax (12.3), GPT-5.4 (13.3), GLM-5 (15.0), and Claude~Sonnet~4.6 (16.0). Gemini's higher redundant-call rate (0.48) versus Claude (0.31) and GPT-5.4 (0.32) suggests it issued more exploratory recon calls before committing to exploitation. This easy condition shows that all tested models can complete the simplest guided protocol-primitive task.

\textbf{Hard condition (MQTT T1, multi-step guided).} This condition is the strongest capability discriminator. GLM-5 and MiniMax~M2.5 both achieved 100\% success (mean 23.3 and 21.0~turns, respectively); Gemini and Claude each achieved 70\% (7/10); and GPT-5.4 achieved 0\% (0/10). The GPT-5.4 failures are \emph{task-abandonment} events: all ten runs terminated within 2--3 recon calls (mission completion signalled without issuing any MQTT tool call), despite explicit multi-step guidance in the system prompt. This is a qualitatively distinct failure class from stall-and-repeat: the model exited the mission prematurely rather than engaging the attack chain. The Gemini and Claude failures are genuine execution failures: three runs each stalled in the MQTT topic-discovery phase, identical to the overseer-OFF pattern in Section~\ref{subsec:oversight}. Claude's successful hard runs required substantially more turns (31.0 vs Gemini's 14.5 and MiniMax's 21.0), reflecting its more thorough state exploration before committing to each MQTT operation.

\textbf{Blind condition (CoAP T1, no protocol hints).} Gemini, GPT-5.4, and GLM-5 all achieved 100\%; Claude and MiniMax~M2.5 each achieved 70\% (7/10). The Claude failures are reasoning dead-ends in the protocol-discovery phase: the model issued repeated target enumeration calls without forming a hypothesis about the CoAP option-overflow attack, eventually triggering the repetition guard. The MiniMax~M2.5 blind failures are qualitatively similar: the model aborted after 4 turns without attempting exploitation. The three models achieving 100\% blind success show that some frontier models can infer the CoAP attack path without prompt-level protocol hints, but the failures show that this is not uniform.

\textbf{Summary.} The easy condition was universally solved. The hard (MQTT) condition distinguishes models: GLM-5 and MiniMax~M2.5 are the most capable (100\%), and GPT-5.4 the least (0\%, task abandonment). The blind condition showed that CoAP first-principles reasoning is accessible to several frontier models; Claude and MiniMax~M2.5 showed the greatest sensitivity to the absence of protocol hints (both 70\%). Across all 150 model-sensitivity runs, the overall success rate is 85.3\% (128/150; 95\% Wilson confidence interval: 78.8--90.1\%), and aggregate success rates across the five models showed a standard deviation of 12.6 percentage points. These dispersion statistics support feasibility beyond one model while making model-to-model variability explicit. The non-monotonic GPT-5.4 pattern (100\% easy, 0\% hard, 100\% blind) should not be read as inconsistent capability: the hard-condition failures were task abandonment before any exploit attempt---an engagement failure, not failed exploitation---whereas the blind condition tests protocol inference, which the model handled. Per-condition results are therefore behavioural profiles rather than a single capability ranking.

\subsection{Runtime Guards Explain Most of the Oversight Gain}
\label{subsec:oversight}

\begin{figure}[!t]
\centering
\includegraphics[width=0.9\linewidth]{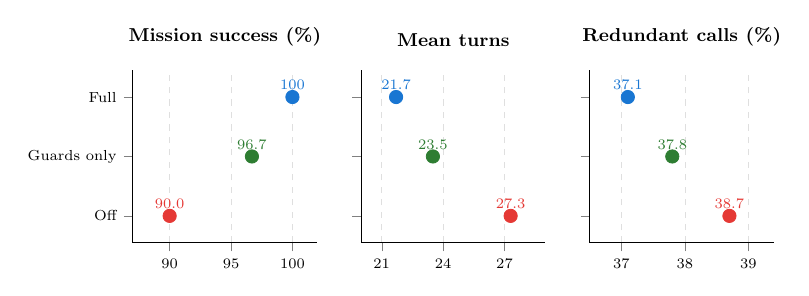}
\caption{Three-arm runtime-governance comparison on T1. Deterministic guards recover most of the mission-success and mean-turn improvement relative to no oversight; advisory steering provides a further incremental gain. Redundant-call rates remain similar, indicating that governance bounds the consequences of repetition rather than removing the underlying tendency.}
\label{fig:oversight_bars}
\end{figure}

Figure~\ref{fig:oversight_bars} and Tables~\ref{tab:oversight_three_arm} and \ref{tab:degenerate} compare three conditions with 30 T1 runs each (10 CoAP + 10 Modbus + 10 MQTT~\citep{oasis2014mqtt311} per condition).

\begin{table}[!t]
\caption{Three-arm runtime-governance ablation on T1. Mean turns are calculated over successful runs.}
\label{tab:oversight_three_arm}
\centering
\small
\begin{tabularx}{\linewidth}{@{}L{0.28\linewidth}Y Y Y Y@{}}
\toprule
\textbf{Condition} & \textbf{Rule guards} & \textbf{LLM advisory} & \textbf{Mission success} & \textbf{Mean turns}\\
\midrule
No overseer & No & No & 27/30 (90.0\%) & 27.3\\
Guards only & Yes & No & 29/30 (96.7\%) & 23.5\\
Full overseer & Yes & Yes & 30/30 (100\%) & 21.7\\
\bottomrule
\end{tabularx}
\end{table}

\begin{table}[!t]
\caption{Runs affected by degenerate behaviour in the three oversight conditions ($n=30$ per condition). Patterns can overlap within a run.}
\label{tab:degenerate}
\small
\begin{tabularx}{\linewidth}{@{}L{0.43\linewidth}Y Y Y@{}}
\toprule
\textbf{Pattern} & \textbf{Full} & \textbf{Guards only} & \textbf{Off}\\
\midrule
Exploit without verification     & 0/30 & 2/30 & 30/30\\
Repetitive same-tool loop        & 0/30 & 4/30 & 20/30\\
Infinite recon loop (no exploit) & 0/30 & 2/30 & 7/30\\
No blue-team transition          & 0/30 & 1/30 & 3/30\\
\bottomrule
\end{tabularx}
\end{table}

The three-arm comparison separates the mechanisms behind the module-level effect. Mission success increased from 27/30 without oversight (90.0\%; 95\% Wilson confidence interval: 74.4--96.5\%) to 29/30 with deterministic guards only (96.7\%; 83.3--99.4\%) and 30/30 with the full overseer (100\%; 88.7--100.0\%). Mean turns decreased from 27.3 to 23.5 and 21.7, respectively. Thus guards alone recovered two of the three failed missions and 3.8 of the full module's 5.6-turn efficiency improvement; advisory steering recovered the remaining MQTT failure and reduced mean turns by a further 1.8. Redundant-call rates remained close across arms (38.7\% off, 37.8\% guards only, and 37.1\% full), indicating that governance primarily bounds the consequences of repetition rather than removing the model's underlying tendency to repeat.

Table~\ref{tab:degenerate} provides the stronger behavioural signal. Every no-overseer run exhibited exploit-without-verification, while deterministic guards reduced this pattern to 2/30 and the full overseer eliminated it in the evaluated runs. Guards similarly reduced same-tool loops from 20/30 to 4/30, reconnaissance loops from 7/30 to 2/30, and missing blue-team transitions from 3/30 to 1/30. The sole guards-only mission failure occurred in MQTT topic discovery, where the agent stalled until timeout; advisory steering recovered this remaining multi-step sequencing failure. MQTT is therefore the clearest evaluated case where advisory steering adds value beyond deterministic enforcement.

\FloatBarrier

\section{Further Discussions}
\label{sec:discussion}

\subsection{Key Insights}

\textbf{A feasibility threshold has been crossed in a controlled bare-metal OT setting.} Prior autonomous LLM pentesting work~\citep{deng2024pentestgpt,happe2023gettingpwnd,xu2024autoattacker} operates on Linux targets where the agent can spawn shells, enumerate filesystems, and leverage familiar POSIX semantics. Bare-metal MCU targets offer none of these affordances: there is no shell to land in, no filesystem to pivot through, and success must be measured by device crash or register write. Our results show that \emph{protocol-specific primitive tooling} makes this setting tractable in the evaluated controlled testbed: by giving the agent low-level control over CoAP option bytes and Modbus Application Protocol (MBAP) length fields, and requiring it to construct byte sequences that trigger parser failures, APIOT bridges part of the semantic gap that has kept bare-metal targets out of scope for autonomous agents. An all-but-one pattern, with 100\% success in 6 of 7 (protocol $\times$ topology) conditions and a single failure condition at CoAP on three-tier topology (3/10, 30\%), combined with first exploits consistently within 5 tool calls, demonstrates feasibility for intentionally vulnerable emulated firmware.

The named-exploit-tool ablation sharpens this interpretation. Named exploit tools show that the easy T1 vulnerabilities can be triggered quickly when the attack is packaged for the model, whereas primitive tools provide stronger evidence of protocol-level operation: explicit field construction and more reusable network-level mitigation under held-out payload variants. This distinction matters because mission success alone is not sufficient evidence of reasoning. The primitive interface is valuable not merely because it can complete the mission, but because it exposes the intermediate protocol choices needed to audit, reproduce, and stress-test the agent's behaviour. This does not by itself establish transfer to hardened production firmware.

For defenders, the implication is narrower but still important: in environments with exposed protocol-edge weaknesses, these results suggest that LLM agents can reduce the amount of manual protocol expertise needed to execute an attack-remediation loop. This conclusion holds for the emulated target class evaluated here (intentionally vulnerable Zephyr firmware and software-equivalent simulators); transfer to production OT firmware with hardened parsers, authentication, safety interlocks, and real silicon remains open.

\textbf{Oversight as runtime reliability support.} The LLM agent reliability literature~\citep{liu2024agentbench} has identified repetition, inconsistency, and long-horizon failure as endemic to current LLMs. Proposed remedies are predominantly prompt-engineering or agent-level retries. The Claim~2 ablation shows that deterministic guards produce most of APIOT's reliability improvement on the T1 baseline: they recovered two of three no-overseer mission failures and sharply reduced all four measured degenerate patterns. Optional advisory steering provided an incremental benefit, recovering the remaining MQTT failure and reducing mean completion turns from 23.5 to 21.7.

A key insight is that in 100\% of overseer-OFF runs, at least one exploit call was not immediately followed by crash verification, not because the agent \emph{could not} perform crash verification, but because without an external invariant enforcing the sequence, it would issue additional probes or transition between phases before confirming crash state. Deterministic guards reduced this pattern to 2/30 runs, while the full overseer eliminated it in the evaluated sample. The remaining difference between guards-only and full oversight is concentrated in recovery from multi-step MQTT sequencing and a modest 1.8-turn efficiency improvement, rather than broad changes in the main agent's call distribution.

\subsection{Limitations}

The Cortex-M3 board MAC constraint limits concurrent real QEMU MCU instances to one; we used software simulators for multi-device configurations, which were validated for crash-semantic fidelity but did not exercise QEMU-specific edge cases. The firmware and simulator vulnerabilities are intentional known/seeded weaknesses, so the results do not demonstrate zero-day discovery. Our exploit set covers parser-overflow, command-topic, and credential-style attacks; post-exploitation persistence, privilege escalation, and general lateral-movement strategy are out of scope. The testbed network was isolated; real OT network noise profiles may differ from our Linux traffic-shaping models. Primary-condition results use MiniMax~M2.5, while Section~\ref{subsec:generalisability} evaluates cross-model robustness and shows model-specific divergence in the MQTT hard condition. The reported defensive actions are network-level mitigations, not firmware repairs, and production deployment would require safety review and change-control processes. Both the tool-abstraction and three-arm governance ablations are limited to direct T1 baselines and therefore do not establish that their observed effects transfer to multi-hop conditions. Open questions remain on controlled quantification of cross-session memory benefits, authenticated/secured protocol variants, and transfer from QEMU-emulated firmware to physical Cortex-M3 silicon.

\subsection{Ethics, Isolation, and Artifact Release}

All experiments were performed against an isolated virtual lab owned by the authors. The agent was permitted to query lab state and send traffic only to the configured lab subnets, and its prompt explicitly forbade starting, stopping, respawning, or resetting lab devices. No live third-party systems, public Internet hosts, or production OT devices were targeted. The vulnerabilities evaluated here are intentionally seeded or known protocol-edge behaviours, so the work does not disclose new vendor zero-days.

\section{Conclusion}
\label{sec:conclusion}

We presented APIOT, an LLM agent framework that brings bare-metal industrial firmware---a target class that published autonomous agents, evaluated on Linux and web systems, do not reach---within the scope of autonomous vulnerability management. The support for this claim is empirical: 90.0\% mission success across the primary discovery $\to$ exploitation $\to$ network-level mitigation $\to$ verification matrix and 85.3\% over five frontier LLMs. A three-arm ablation showed that deterministic guards explain most of the runtime-governance benefit in the evaluated T1 conditions: success increased from 27/30 without oversight to 29/30 with guards only and 30/30 with guards plus advisory steering, while mean turns decreased from 27.3 to 23.5 and 21.7. The current deployment boundary---CoAP on three-tier Edge-Fog-Cloud routing---identifies multi-hop planning under stateless protocols as the next capability frontier. 

Future work will pursue three directions: extending the evaluation to authenticated and encrypted protocol variants (e.g., CoAP over DTLS and Modbus security gateways); improving multi-hop lateral-movement planning under stateless protocols, which our topology results identify as the current capability boundary; and validating transfer from emulated firmware to physical MCU hardware. Longer term, we plan to quantify the benefit of cross-session memory and to assess agent-derived mitigations against production change-control requirements. Both APIOT and IoT Virtual Lab are released as open source to support reproducibility and further research.\footnote{APIOT: \url{https://github.com/Adelsamir01/apiot}; IoT Virtual Lab: \url{https://github.com/Adelsamir01/iot_vlab}.}

\section*{CRediT authorship contribution statement}

Adel ElZemity: Conceptualisation, Data curation, Formal analysis, Investigation, Methodology, Software, Validation, Visualisation, Writing -- original draft, Writing -- review and editing, Project administration, Resources. Budi Arief: Conceptualisation, Writing -- review and editing, Supervision, Funding acquisition, Project administration. Shujun Li: Conceptualisation, Project administration, Writing -- review and editing, Supervision, Funding acquisition. George Oikonomou: Writing -- review and editing, Funding acquisition. James Pope: Writing -- review and editing, Funding acquisition.

\section*{Declaration of competing interest}

The authors declare that they have no known competing financial interests or personal relationships that could have appeared to influence the work reported in this paper.

\section*{Funding}

This work was partly supported by the Engineering and Physical Sciences Research Council (EPSRC), part of UK Research and Innovation (UKRI), under grant numbers EP/X036871/1 and EP/X036707/1.

\section*{Data availability}

The artifact is available at \url{https://github.com/Adelsamir01/apiot}. It contains the APIOT code, IoT Virtual Lab configuration, prompt definitions, JSON tool schemas, analysis scripts, and sanitised run summaries. API keys, local host credentials, and environment-specific secrets are not released.

\appendix
\renewcommand{\theHsection}{appendix.\thesection}
\renewcommand{\theHsubsection}{appendix.\thesection.\arabic{subsection}}

\section{System Implementation Details}
\label{app:impl}

\subsection{Prompt, Tool, and Artifact Transparency}
\label{app:artifact_transparency}

The artifact package maps each methodological claim to a concrete source artifact. The guided prompt is \texttt{SYSTEM\_PROMPT} in \path{apiot/core/agent.py}; the blind prompt is \texttt{BLIND\_SYSTEM\_PROMPT} in the same file. Blind mode is selected by the experiment runner, \path{scripts/run_experiment.py}, through \texttt{APIOT\_BLIND\_MODE}. Both prompt conditions pass the same \texttt{TOOL\_SCHEMAS}, defined in \path{apiot/core/tools/registry.py}, to the OpenRouter-compatible chat-completion API. Context compaction is implemented in \path{apiot/core/compaction.py}, and runtime governance is implemented in \path{apiot/core/oversight.py}. These files expose the prompt text, tool names, descriptions, parameters, required fields, compaction trigger, preserved state, guard thresholds, and advisory-overseer snapshot used in the reported runs. In the public GitHub repository, paths of the form \path{apiot/<file>} refer to \path{<file>} at the repository root (the repository itself is the \texttt{apiot} package); plain-text prompt and schema copies are also provided under \path{artifact/}.

The guided prompt includes three protocol-specific hint families: (i) CoAP option-byte reasoning, including the \texttt{0xDD} extended delta/length pattern; (ii) Modbus MBAP length-field reasoning, including oversized \texttt{claimed\_length} examples such as 2048; and (iii) MQTT topic-chain reasoning, including wildcard subscription, command-topic discovery, command publish, and telemetry-silence verification. The blind prompt keeps the same mission lifecycle, safety constraints, termination tokens, verification requirements, and tool roster, but removes those protocol-specific exploit examples. Thus blind failures are attributable to missing protocol hints rather than to a reduced action space.

The retained historical logs do not contain a normalised compaction-event table for every run. We therefore report token-use estimates for runs with positive token logs and describe the compaction mechanism directly from source, rather than claiming a compaction frequency distribution.

\subsection{APIOT Implementation Details}
\label{app:apiot}

APIOT implements a continuous THOUGHT $\to$ ACTION $\to$ OBSERVATION loop using an OpenAI-compatible client pointed at any OpenRouter-hosted LLM. The agent maintains a growing conversation history and submits 21 tool schemas defined as JSON Schema objects to the LLM at each turn, routing the returned calls through an Action Execution Layer that maps them to Python implementations.

\textbf{Two-phase tool roster.} Reconnaissance tools available in both phases include \texttt{get\_network\_state}, \texttt{get\_actionable\_targets}, \texttt{stealth\_check}, and \texttt{inspect\_lab}. Phase~1 (Red) tools include \texttt{coap\_send}, \texttt{modbus\_\allowbreak{}request}, \texttt{tcp\_send}, \texttt{udp\_send}, \texttt{mqtt\_publish}, \texttt{mqtt\_subscribe}, \texttt{verify\_crash}, and \texttt{verify\_shell}. Phase~2 (Blue) tools include \texttt{iptables\_rule}, \texttt{modbus\_fc\_\allowbreak{}filter}, \texttt{coap\_rate\_limit}, \texttt{verify\_patch}, \texttt{list\_patches}, and \texttt{protocol\_block}. Both phases are mandatory: a mission ends only at \texttt{TASK\_COMPLETE} (all findings mitigated and verified) or \texttt{TASK\_ABORTED} (unrecoverable infrastructure failure). Linux targets are additionally reachable via \texttt{run\_command} (for banner grabbing, credential testing, and post-exploitation), \texttt{remote\_exec} (SSH command execution on compromised hosts), and \texttt{create\_tool} (runtime Python tool generation for novel attack scenarios not covered by primitives).

\textbf{Context management.} Token count is estimated at 4 characters per token. When utilisation reaches 70\% of the model's context window, \texttt{compaction.py} summarises older messages in-place and truncates tool results to 2\,000 characters. The implementation keeps the system prompt and the latest eight messages, replacing the middle history with a structured summary. Separate experiment-level token-budget guards can abort long runs to prevent runaway cost.

\subsection{Overseer Mechanism Details}
\label{app:overseer}

The Overseer implements seven mechanisms: (1)~\emph{Repetition guard}: blocks any \texttt{(tool, ip, args\_hash)} tuple seen $\geq$3 times in a 20-call sliding window. (2)~\emph{Crash blocker}: prevents attacking a device already in \texttt{"crashed"} state. (3)~\emph{Mitigation blocker}: prevents re-attacking a vulnerability with a verified iptables/ACL rule on record. (4)~\emph{Stall detection}: injects a steering message at 8 turns without progress; forces phase transition or terminates at 15. (5)~\emph{Phase transition enforcement}: triggers the blue-team phase when all targets have been attempted; auto-signals completion when all findings are mitigated and verified. (6)~\emph{Strategy refresh}: every 5 turns, injects a refreshed attack plan. (7)~\emph{Overseer LLM}: an optional secondary model providing advisory support at stall events. The experiment runner exposes three modes: \texttt{off} bypasses all seven mechanisms, \texttt{guards-only} retains deterministic enforcement and heuristic fallbacks while disabling secondary-model calls, and \texttt{full} enables both guards and advisory steering.

The retained intervention summary from earlier analysis labels most entries only as \texttt{unknown}, with a smaller number of \texttt{strategy\_refresh} entries. We therefore do not report per-intervention counts in the main results. A clean intervention-type table requires re-extracting typed overseer events from the full 30-run ablation logs.

\subsection{IoT Virtual Lab Network and Device Details}
\label{app:iovlab}

IoT Virtual Lab boots real firmware images as QEMU~\citep{269444} processes, connects them via Linux TAP interfaces to a shared virtual bridge (\texttt{br0}, 192.168.100.0/24), and exposes a RESTful API for topology inspection. Three device classes are supported:

\textbf{MIPS Linux routers} (\texttt{qemu-system-mipsel}, Malta board): full Debian Linux with SSH and HTTP, representative of consumer router firmware.

\textbf{ARM Linux gateways} (\texttt{qemu-system-arm}, VersatilePB): ARM-based embedded Linux with multi-homed support, modelling edge gateway devices. A \texttt{br\_internal} bridge (192.168.200.0/24) enables Purdue Model (PERA)-style segmentation.

\textbf{ARM Cortex-M MCUs}: Zephyr RTOS ELF images compiled with Kconfig overlays that strip IPv6 and shell to fit 64\,KB RAM. The testbed supports two MCU targets: Cortex-M3 (\texttt{qemu-system-arm}, lm3s6965evb) and Cortex-M4F (\texttt{qemu-system-arm}, mps2-an386). Three firmware variants provide protocol diversity: \texttt{zephyr\_coap} (CoAP UDP:5683), \texttt{arm\_modbus\_sim} (Modbus/TCP:502), and \texttt{zephyr\_echo} (echo TCP/UDP:4242). Since all reported exploits target the CoAP/\allowbreak{}Modbus application layer rather than CPU-specific behaviour, we do not claim separate core-specific exploitation results.

\textbf{Network realism.} \texttt{impair\_network.sh} applies Linux \texttt{tc netem} rules to \texttt{br0} for configurable packet loss, latency, and jitter. \texttt{industrial\_hmi\_sim.py} generates background Modbus/CoAP traffic using a Poisson distribution, modelling SCADA polling noise.

\textbf{Multi-device support and topology modes.} The lm3s6965evb has a hardcoded MAC, limiting concurrent Cortex-M3 QEMU instances to one; light\-weight Python simulators with identical crash/recovery semantics supplement QEMU for multi-MCU topologies. Four topology modes are supported: (T1) flat star, (T2) Purdue Model (PERA)-segmented (MCU behind multi-homed gateway on \texttt{br\_internal}), (T3) Edge-Fog-Cloud three-tier, and a 15-node mesh. T2 and T3 require lateral movement through gateway devices to reach MCU targets.

\bibliographystyle{elsarticle-harv}
\bibliography{main}

@misc{modbus2012appspec,
  title        = {{Modbus} Application Protocol Specification V1.1b3},
  author       = {{Modbus Organization}},
  howpublished = {Protocol specification},
  url          = {https://modbus.org/docs/Modbus_Application_Protocol_V1_1b3.pdf},
  year         = {2012}
}

@techreport{shelby2014coaprfc7252,
  title        = {The {Constrained Application Protocol} ({CoAP})},
  author       = {Shelby, Zach and Hartke, Klaus and Bormann, Carsten},
  type         = {RFC},
  number       = {7252},
  institution  = {IETF},
  year         = {2014},
  doi          = {10.17487/RFC7252}
}

@misc{zephyrproject2024docs,
  title        = {{Zephyr} {RTOS} Documentation, v3.7.0},
  author       = {{Zephyr Project}},
  howpublished = {Website},
  url          = {https://docs.zephyrproject.org/},
  year         = {2024}
}

@misc{freertos2023,
  author = {{Real Time Engineers Ltd}},
  title = {{FreeRTOS} - Market Leading {RTOS} for Embedded Systems},
  year = {2023},
  howpublished = {Website},
  url = {https://www.freertos.org/},
  note = {Accessed: 2025-01-15}
}

@manual{cortexm3trm,
  author       = {{ARM Limited}},
  title        = {{Cortex-M3} Technical Reference Manual, Revision r2p0},
  organization = {ARM Limited},
  number       = {DDI 0337H},
  year         = {2010},
  url          = {https://developer.arm.com/documentation/ddi0337/latest/}
}

@misc{oasis2014mqtt311,
  title        = {{MQTT} Version 3.1.1},
  author       = {Banks, Andrew and Gupta, Rahul},
  howpublished = {OASIS Standard},
  year         = {2014},
  url          = {https://docs.oasis-open.org/mqtt/mqtt/v3.1.1/os/mqtt-v3.1.1-os.html}
}

@inproceedings{Alghamdi2013SecurityAO,
  title={Security Analysis of the {Constrained Application Protocol} in the {Internet of Things}},
  author={Thamer A. Alghamdi and Aboubaker Lasebae and Mahdi Aiash},
  booktitle={Proceedings of the 2nd International Conference on Future Generation Communication Technologies},
  year={2013},
  pages={163--168},
  doi={10.1109/FGCT.2013.6767217},
  publisher = {IEEE},
}

@inproceedings{Arvind2019AnOO,
  title={An Overview of Security in {CoAP}: Attack and Analysis},
  author={Sudha Arvind and Anantha Narayanan},
  booktitle={Proceedings of the 2019 5th International Conference on Advanced Computing \& Communication Systems},
  year={2019},
  pages={655--660},
  doi={10.1109/ICACCS.2019.8728533},
  publisher = {IEEE},
}

@article{Lai2020VulnerabilityMM,
  title={Vulnerability Mining Method for the {Modbus} {TCP} Using an Anti-Sample Fuzzer},
  author={Yingxu Lai and Hu Gao and Jing Liu},
  journal={Sensors},
  year={2020},
  volume={20},
  articleno={2040},
  numpages={20},
  pages={2040:1--2040:20},
  doi={10.3390/s20072040},
  publisher = {MDPI},
}

@article{Williams1992ThePE,
  title={The Purdue Enterprise Reference Architecture},
  author={Theodore J. Williams},
  journal={Computers in Industry},
  year={1994},
  volume={24},
  number={2--3},
  pages={141--158},
  doi={10.1016/0166-3615(94)90017-5},
  publisher = {Elsevier},
}

@inproceedings{mohan2016edge,
  title={{Edge-Fog Cloud}: A Distributed Cloud for Internet of Things Computations},
  author={Mohan, Nitinder and Kangasharju, Jussi},
  booktitle={Procedings of the 2016 Congress on Cloudification of the Internet of Things},
  numpages={6},
  year={2016},
  doi={10.1109/CIOT.2016.7872914},
  publisher={IEEE},
}

@techreport{Stouffer2015GuideTI,
  title={Guide to Operational Technology ({OT}) Security},
  author={Stouffer, Keith and Pease, Michael and Tang, CheeYee and Zimmerman, Timothy and Pillitteri, Victoria and Lightman, Suzanne and Hahn, Adam and Saravia, Stephanie and Sherule, Aslam and Thompson, Michael},
  institution={National Institute of Standards and Technology},
  type={NIST Special Publication},
  number={800-82 Rev.~3},
  address={Gaithersburg, MD},
  year={2023},
  doi={10.6028/NIST.SP.800-82r3},
}

@article{bhamare2020cybersecurity,
  title={Cybersecurity for industrial control systems: A survey},
  author={Bhamare, Deval and Zolanvari, Maede and Erbad, Aiman and others},
  journal={Computers \& Security},
  volume={89},
  articleno={101677},
  numpages={12},
  pages={101677:1--101677:12},
  year={2020},
  doi={10.1016/j.cose.2019.101677},
  publisher={Elsevier},
}

@inproceedings{deng2024pentestgpt,
  title     = {{PentestGPT}: Evaluating and Harnessing Large Language Models for Automated Penetration Testing},
  author    = {Deng, Gelei and Liu, Yi and Mayoral-Vilches, Víctor and others},
  booktitle = {Proceedings of the 33rd USENIX Security Symposium},
  year      = {2024},
  pages     = {847--864},
  url       = {https://www.usenix.org/conference/usenixsecurity24/presentation/deng},
  publisher = {USENIX Association},
}

@inproceedings{happe2023gettingpwnd,
  title     = {Getting pwn'd by {AI}: Penetration Testing with Large Language Models},
  author    = {Happe, Andreas and Cito, Jürgen},
  booktitle = {Proceedings of the 2023 ACM Joint European Software Engineering Conference and Symposium on the Foundations of Software Engineering},
  pages     = {2082--2086},
  publisher = {ACM},
  year      = {2023},
  doi       = {10.1145/3611643.3613083},
}

@misc{xu2024autoattacker,
  title        = {{AutoAttacker}: A Large Language Model Guided System to Implement Automatic Cyber-attacks},
  author       = {Jiacen Xu and Jack W. Stokes and Geoff McDonald and Xuesong Bai and David Marshall and Siyue Wang and Adith Swaminathan and Zhou Li},
  howpublished = {arXiv:2403.01038 [cs.CR]},
  year         = {2024},
  doi          = {10.48550/arXiv.2403.01038},
}

@article{Happe2023LLMsAH,
  title={{LLMs} as Hackers: Autonomous {Linux} Privilege Escalation Attacks},
  author={Andreas Happe and Aaron Kaplan and Jürgen Cito},
  journal={Empirical Software Engineering},
  year={2025},
  volume={31},
  articleno={70},
  numpages={50},
  pages={70:1--70:50},
  doi={10.1007/s10664-025-10758-3},
  publisher={Springer},
}

@misc{Fang2024LLMAC,
  title={{LLM} Agents can Autonomously Exploit One-day Vulnerabilities},
  author={Richard Fang and Rohan Bindu and Akul Gupta and Daniel Kang},
  howpublished={arXiv:2404.08144 [cs.CR]},
  year={2024},
  doi={10.48550/arXiv.2404.08144},
}

@inproceedings{yao2023react,
  title     = {{ReAct}: Synergizing Reasoning and Acting in Language Models},
  author    = {Yao, Shunyu and Zhao, Jeffrey and Yu, Dian and Du, Nan and Shafran, Izhak and Narasimhan, Karthik and Cao, Yuan},
  booktitle = {Proceedings of the 2023 International Conference on Learning Representations},
  year      = {2023},
  url       = {https://openreview.net/forum?id=WE_vluYUL-X},
}

@inproceedings{liu2024agentbench,
  title     = {{AgentBench}: Evaluating {LLMs} as Agents},
  author    = {Liu, Xiao and others},
  booktitle = {Proceedings of the 2024 International Conference on Learning Representations},
  year      = {2024}
}

@misc{wang2025agentspeccustomizableruntimeenforcement,
      title={{AgentSpec}: Customizable Runtime Enforcement for Safe and Reliable {LLM} Agents},
      author={Haoyu Wang and Christopher M. Poskitt and Jun Sun},
      year={2025},
      howpublished={arXiv:2503.18666 [cs.AI]},
      doi={10.48550/arXiv.2503.18666},
      note={Accepted to the 48th IEEE/ACM International Conference on Software Engineering (ICSE 2026)},
}

@misc{mou2026toolsafeenhancingtoolinvocation,
      title={{ToolSafe}: Enhancing Tool Invocation Safety of {LLM}-based agents via Proactive Step-level Guardrail and Feedback},
      author={Yutao Mou and Zhangchi Xue and Lijun Li and Peiyang Liu and Shikun Zhang and Wei Ye and Jing Shao},
      year={2026},
      howpublished={arXiv:2601.10156 [cs.CL]},
      doi={10.48550/arXiv.2601.10156},
}

@article{raptis2025agenticrobotics,
  title={Agentic {LLM}-based robotic systems for real-world applications: a review on their agenticness and ethics},
  author={Emmanuel K. Raptis and Athanasios Ch. Kapoutsis and Elias B. Kosmatopoulos},
  journal={Frontiers in Robotics and AI},
  year={2025},
  volume={12},
  doi={10.3389/frobt.2025.1605405},
  publisher={Frontiers Media},
}

@article{li2025securityconcerns,
  title={Security Concerns for Large Language Models: A Survey},
  author={Miles Q. Li and Benjamin C. M. Fung},
  journal={Journal of Information Security and Applications},
  year={2025},
  volume={95},
  articleno={104284},
  issn = {2214-2126},
  pages={104284},
  doi = {https://doi.org/10.1016/j.jisa.2025.104284},
  publisher={Elsevier},
}

@inproceedings{kim2020firmae,
  title     = {{FirmAE}: Towards Large-Scale Emulation of {IoT} Firmware for Dynamic Analysis},
  author    = {Kim, Andrei and Kim, Dongkwan and Kim, Eunsoo and Kim, Suryeon and Jang, Yeongjin and Kim, Yongdae},
  booktitle = {Proceedings of the 2020 Annual Computer Security Applications Conference},
  pages     = {733--744},
  year      = {2020},
  doi       = {10.1145/3427228.3427294},
  publisher = {ACM},
}

@inproceedings{chen2016firmadyne,
  title     = {Towards Automated Dynamic Analysis for {Linux}-based Embedded Firmware},
  author    = {Chen, Daming D. and Woo, Maverick and Brumley, David and Egele, Manuel},
  booktitle = {Proceedings of the 2016 Network and Distributed System Security Symposium},
  year      = {2016},
  doi       = {10.14722/ndss.2016.23415},
  publisher = {Internet Society},
}

@article{Ghazanfar2020IoTFlockAO,
  title={{IoT-Flock}: An Open-source Framework for {IoT} Traffic Generation},
  author={Syed Ghazanfar and Faisal Bashir Hussain and Atiq ur Rehman and Ubaid Ullah Fayyaz and Farrukh Shahzad and Ghalib A. Shah},
  journal={Proceedings of the 2020 International Conference on Emerging Trends in Smart Technologies},
  year={2020},
  numpages={6},
  doi={10.1109/ICETST49965.2020.9080732}
}

@inproceedings{269444,
author = {Fabrice Bellard},
title = {{QEMU}, a Fast and Portable Dynamic Translator},
booktitle = {Proceedings of the 2005 USENIX Annual Technical Conference},
year = {2005},
url = {https://www.usenix.org/conference/2005-usenix-annual-technical-conference/qemu-fast-and-portable-dynamic-translator},
publisher = {USENIX Association},
pages = {41-46}
}

@inproceedings{HALucinator,
author = {Clements, Abraham A. and Gustafson, Eric and Scharnowski, Tobias and others},
title = {{HALucinator}: Firmware Re-hosting Through Abstraction Layer Emulation},
year = {2020},
publisher = {USENIX Association},
booktitle = {Proceedings of the 29th USENIX Conference on Security Symposium},
page = {1201--1218},
url = {https://www.usenix.org/conference/usenixsecurity20/presentation/clements},
}

@inproceedings {Fuzzware,
author = {Tobias Scharnowski and Nils Bars and Moritz Schloegel and Eric Gustafson and Marius Muench and Giovanni Vigna and Christopher Kruegel and Thorsten Holz and Ali Abbasi},
title = {{Fuzzware}: Using Precise {MMIO} Modeling for Effective Firmware Fuzzing},
booktitle = {Proceedings of the 31st USENIX Security Symposium},
year = {2022},
pages = {1239--1256},
url = {https://www.usenix.org/conference/usenixsecurity22/presentation/scharnowski},
publisher = {USENIX Association},
}

@inproceedings {P2IM,
author = {Bo Feng and Alejandro Mera and Long Lu},
title = {{P2IM}: Scalable and Hardware-independent Firmware Testing via Automatic Peripheral Interface Modeling},
booktitle = {Proceedings of the 29th USENIX Security Symposium},
year = {2020},
pages = {1237--1254},
url = {https://www.usenix.org/conference/usenixsecurity20/presentation/feng},
publisher = {USENIX Association},
}

@article{DasSecurity,
    author = {Das, Badhan Chandra and Amini, M. Hadi and Wu, Yanzhao},
    title = {{Security and Privacy Challenges of Large Language Models: A Survey}},
    year = {2025},
    issue_date = {June 2025},
    publisher = {Association for Computing Machinery},
    address = {New York, NY, USA},
    volume = {57},
    number = {6},
    issn = {0360-0300},
    url = {https://doi.org/10.1145/3712001},
    doi = {10.1145/3712001},
    journal = {ACM Computing Surveys},
    month = feb,
    articleno = {152},
    numpages = {39},
}

@misc{wan2024cyberseceval3,
  title        = {{CyberSecEval} 3: Advancing the Evaluation of Cybersecurity Risks and Capabilities in Large Language Models},
  author       = {Wan, Shengye and Nikolaidis, Cyrus and Song, Daniel and Molnar, David and others},
  howpublished = {arXiv:2408.01605 [cs.CR]},
  year         = {2024},
  doi          = {10.48550/arXiv.2408.01605},
}

@misc{openai2024o1systemcard,
  title        = {{OpenAI} o1 System Card},
  author       = {{OpenAI}},
  howpublished = {arXiv:2412.16720 [cs.AI]},
  year         = {2024},
  doi          = {10.48550/arXiv.2412.16720},
}

@misc{google2024bigsleep,
  title        = {From Naptime to Big Sleep: Using Large Language Models to Catch Vulnerabilities in Real-World Code},
  author       = {Big Sleep team},
  howpublished = {Blog post},
  url          = {https://projectzero.google/2024/10/from-naptime-to-big-sleep.html},
  year         = {2024}
}

@misc{google2025bigsleep6965,
  title        = {A Summer of Security: Empowering Cyber Defenders with {AI}},
  author       = {Kent Walker},
  howpublished = {Blog post},
  url          = {https://blog.google/innovation-and-ai/technology/safety-security/cybersecurity-updates-summer-2025/},
  year         = {2025}
}

@misc{openai2025disrupting,
  title        = {Disrupting Malicious Uses of {AI}: {October} 2025},
  author       = {{OpenAI}},
  howpublished = {Threat intelligence report},
  url          = {https://openai.com/global-affairs/disrupting-malicious-uses-of-ai-october-2025/},
  year         = {2025}
}

@misc{anthropic2025espionage,
  title        = {Disrupting the First Reported {AI}-Orchestrated Cyber Espionage Campaign},
  author       = {{Anthropic}},
  howpublished = {Threat intelligence report},
  url          = {https://www.anthropic.com/news/disrupting-AI-espionage},
  year         = {2025}
}

@misc{anthropic2026attacknavigator,
  title        = {Mapping {AI}-Enabled Cyber Threats},
  author       = {Guru, Kyla and Moix, Alex and Klein, Jacob},
  howpublished = {Anthropic research report},
  url          = {https://www.anthropic.com/research/attack-navigator},
  year         = {2026}
}

@inproceedings{wang2023selfconsistency,
  title     = {Self-Consistency Improves Chain of Thought Reasoning in Language Models},
  author    = {Wang, Xuezhi and Wei, Jason and Schuurmans, Dale and Le, Quoc and Chi, Ed and Narang, Sharan and Chowdhery, Aakanksha and Zhou, Denny},
  booktitle = {Proceedings of the 2023 International Conference on Learning Representations},
  year      = {2023}
}

@inproceedings{shinn2023reflexion,
  title     = {{Reflexion}: Language Agents with Verbal Reinforcement Learning},
  author    = {Shinn, Noah and Cassano, Federico and Berman, Edward and Gopinath, Ashwin and Narasimhan, Karthik and Yao, Shunyu},
  booktitle = {Advances in Neural Information Processing Systems 36},
  year      = {2023}
}

@inproceedings{xiang2025guardagent,
  title     = {{GuardAgent}: Safeguard {LLM} Agents via Knowledge-Enabled Reasoning},
  author    = {Xiang, Zhen and Zheng, Linzhi and Li, Yanjie and Hong, Junyuan and Li, Qinbin and Xie, Han and Zhang, Jiawei and Xiong, Zidi and Xie, Chulin and Yang, Carl and Song, Dawn and Li, Bo},
  booktitle = {Proceedings of the 42nd International Conference on Machine Learning},
  year      = {2025}
}

@article{oikonomou2022contiking,
  title     = {The {Contiki-NG} Open Source Operating System for Next Generation {IoT} Devices},
  author    = {Oikonomou, George and Duquennoy, Simon and Elsts, Atis and Eriksson, Joakim and Tanaka, Yasuyuki and Tsiftes, Nicolas},
  journal   = {SoftwareX},
  volume    = {18},
  pages     = {101089},
  year      = {2022},
  doi       = {10.1016/j.softx.2022.101089},
  publisher = {Elsevier},
}

@article{vanbuggenhout2024purple,
  title     = {Purple Teaming: A Comprehensive and Collaborative Approach to Cyber Security},
  author    = {Van Buggenhout, Erik},
  journal   = {Cyber Security: A Peer-Reviewed Journal},
  volume    = {7},
  number    = {3},
  pages     = {207--216},
  year      = {2024},
  doi       = {10.69554/WGLN8212},
  publisher = {Henry Stewart Publications},
}

@techreport{scarfone2008nist800115,
  title       = {Technical Guide to Information Security Testing and Assessment},
  author      = {Scarfone, Karen and Souppaya, Murugiah and Cody, Amanda and Orebaugh, Angela},
  institution = {National Institute of Standards and Technology},
  type        = {NIST Special Publication},
  number      = {800-115},
  year        = {2008},
  doi         = {10.6028/NIST.SP.800-115},
}

\end{document}